\documentclass[a4paper,12pt]{article}
\usepackage{graphicx,amssymb}

\newcommand{\bea}{\begin{eqnarray}}
\newcommand{\ena}{\end{eqnarray}}
\newcommand{\vs}[1]{\vspace{#1 mm}}
\newcommand{\hs}[1]{\hspace{#1 mm}}
\renewcommand{\a}{\alpha}
\renewcommand{\b}{\beta}
\renewcommand{\c}{\gamma}

\newcommand{\e}{\epsilon}
\newcommand{\s}{\sigma}

\def\bbox{{\,\lower0.9pt\vbox{\hrule \hbox{\vrule height 0.2 cm
\hskip 0.2 cm \vrule height 0.2 cm}\hrule}\,}}
\newcommand{\dsl}{\pa \kern-0.5em /}

\newcommand{\shalf}{\frac{1}{2}}
\newcommand{\pa}{\partial}
\newcommand{\nn}{\nonumber\\}
\newcommand{\p}[1]{(\ref{#1})}

\begin{document}

\renewcommand{\thefootnote}{\fnsymbol{footnote}}
\begin{titlepage}

\setcounter{page}{0}
\begin{flushright}
OU-HET 434 \\
USTC-ICTS-03-4 \\
hep-th/0303238
\end{flushright}

\vs{10}
\begin{center}
{\Large\bf Accelerating Cosmologies from S-Branes}
\vs{20}

{\large
Nobuyoshi Ohta\footnote{e-mail address: ohta@phys.sci.osaka-u.ac.jp}}\\
\vs{10}
{\em Department of Physics, Osaka University,
Toyonaka, Osaka 560-0043, Japan}\\
and \\
{\em Interdisciplinary Center for Theoretical Study,
University of Science and Technology of China,
Hefei, Anhui 230026, P. R. China}
\end{center}
\vs{15}
\centerline{{\bf{Abstract}}}
\vs{5}

We point out that the recently proposed model of a flat 4-dimensional
universe with accelerated expansion in string/M-theory is a special case
of time-dependent solutions that the author found under the name of
``S-branes.''
We also show that similar accelerating models can be obtained from S-branes
if the internal space is chosen to be hyperbolic or flat spaces.

\vs{10}

\end{titlepage}
\newpage
\renewcommand{\thefootnote}{\arabic{footnote}}
\setcounter{footnote}{0}
\setcounter{page}{2}

In a recent paper~\cite{TW}, Townsend and Wohlfarth proposed a solution of
$(4+n)$-dimensional vacuum Einstein equations in string/M-theory with compact
hyperbolic internal space which exhibits accelerated expansion (hereafter
referred to as accelerating solution). This is an interesting cosmological
model since astronomical observations show that the universe is
not only expanding but also is undergoing accelerated
expansion~\cite{AR}. Recent measurements of the cosmic microwave background
seem to further support the accelerated expansion in an inflationary
epoch~\cite{LB}. Related discussions can be found in refs.~\cite{HW}-\cite{GH}.

On the other hand, an interesting class of time-dependent solutions have been
found in the supergravity theories in higher dimensions, the low-energy
effective theories of superstring/M-theory. These are the
spacelike brane solutions (S-branes) which were proposed in connection with
tachyon condensations and dS/CFT correspondence~\cite{GS}-\cite{NO1}, but the
present interest is concerned with their property as time-dependent solutions.
In particular, the analysis in ref.~\cite{NO1} is quite general to discuss
time-dependent solutions, and one may wonder if there is any connection
between the S-brane and above solutions.

At first sight, it may appear that there is no connection since the above
accelerating solution is the one to the vacuum Einstein equations whereas
S-branes are a class of solutions with background antisymmetric tensors.
It is true for the solutions in refs.~\cite{GS}-\cite{V} since these
necessarily involve nonzero field strengths. However, we would like to point
out that our solutions in ref.~\cite{NO1} are sufficiently general to cover
the accelerating solution, which is actually a special case of the
time-dependent solutions that the present author derived.
We show that our solutions~\cite{NO1} reduce to the accelerating solution
if we put the field strength to zero and choose constants appropriately.
In addition, we show that more general S-brane solutions exhibit similar
accelerated expansion if we choose the compact internal space to be
hyperbolic. It turns out that actually the internal flat space is also
allowed for accelerating solution, thus providing wider class of solutions
appropriate for cosmology. Following the usual convention, we use S$q$-branes
for those with $(q+1)$-dimensional Euclidean world-volume.

The solution in ref.~\cite{TW} is the one for $(4+n)$-dimensional vacuum
Einstein equations
\bea
ds^2 = e^{3nt/(n-1)} K^{-n/(n-1)} ds_E^2 + e^{-6t/(n-1)} K^{2/(n-1)}
d\Sigma_n^2,
\label{exsol1}
\ena
where $n$ is the dimension of the internal hyperbolic space, which is
compactified, and
\bea
ds_E^2 = -S^6 dt^2 + S^2 d{\bf x}^2,
\label{exsol2}
\ena
describes the 4-dimensional spacetime with
\bea
S(t) &=& e^{-(n+2)t/2(n-1)} K^{n/2(n-1)}, \nn
K(t) &=& \frac{\sqrt{3(n+2)/n}}{(n-1)\sinh(\sqrt{3(n+2)/n}\; |t|)}.
\label{factor}
\ena
If we take the time coordinate $\eta$ defined by
\bea
d\eta= S^3(t) dt,
\label{time}
\ena
the metric~\p{exsol2} describes a flat homogeneous isotropic universe with
scale factor $S$. The condition for expanding 4-dimensional universe is that
\bea
\frac{dS}{d\eta}>0.
\label{cond1}
\ena
Accelerated expansion is obtained if, in addition,
\bea
\frac{d^2S}{d\eta^2}>0.
\label{cond2}
\ena
It has been shown that these can be satisfied for $n=7$ and for certain period
of negative $t$ which is the period that our universe is evolving ($t<0$ and
$t>0$ are two disjoint possible universes)~\cite{TW}.

We are now going to show that our solutions in ref.~\cite{NO1} reduce to
this if we set the field strength to zero. Our action consists of gravity
coupled to a dilaton $\phi$ and $m$ different $n_A$-form field strengths
in arbitrary dimensions $d$. It describes the bosonic part of the $d=11$ or
$d=10$ supergravities if we choose the parameters suitably.

The solutions are given by
\bea
\label{oursol}
ds_d^2 &=& \prod_A [\cosh\tilde c_A (t-t_A)]^{2 \frac{q_A+1}{\Delta_A}}
\Bigg[ e^{2c_0 t+2c_0'} \left\{ - e^{2ng(t)} dt^2
+ e^{2g(t)}d\Sigma_{n,\s}^2\right\} \nn
&& \hs{20} +\; \sum_{\a=1}^{p} \prod_A [\cosh\tilde c_A(t-t_A)]^{- 2
\frac{\c_A^{(\a)}}{\Delta_A}} e^{2 \tilde c_\a t+2c_\a'} dx_\a^2\Bigg], \\
E_A &=& \sqrt{\frac{2(d-2)}{\Delta_A}}\frac{e^{\tilde c_A(t-t_A)
-\e_Aa_A c_\phi'/2+\sum_{\a\in q_A} c_\a'}}{\cosh \tilde c_A(t-t_A)},\quad
\tilde c_A = \sum_{\a\in q_A} c_\a-\frac{1}{2} c_\phi \e_A a_A, \nn
\phi &=& \sum_{A} \frac{(d-2)\e_A a_A}{\Delta_A} \ln \cosh\tilde c_A(t-t_A)
+\tilde c_\phi t + c_\phi',
\label{oursol1}
\ena
where $d=p+n+1$, $A$ denotes the kinds of $q_A$-branes, the time derivatives
of $E_A$ give the values of the field strengths of antisymmetric tensors,
$a_A$ is the parameter for the coupling of dilaton and forms,
and $\e_A= +1 (-1)$ corresponds to electric (magnetic) fields.
The coordinates $x_\a, (\a=1,\ldots, p)$ parametrize the $p$-dimensional
world-volume directions and the remaining coordinates of the $d$-dimensional
spacetime are the time $t$ and coordinates on compact $n$-dimensional spherical
($\s=+1$), flat ($\s=0$) or hyperbolic ($\s=-1$) spaces, whose line elements
are $d\Sigma_{n,\s}^2$.
We have also defined
\bea
\label{res2}
&& \Delta_A = (q_A + 1) (d-q_A-3) + \shalf a_A^2 (d-2), \nn
&& \c_A^{(\a)} = \left\{ \begin{array}{l}
d-2 \\
0
\end{array}
\right.
\hs{5}
{\rm for} \hs{3}
\left\{
\begin{array}{l}
x_\a \hs{3} {\rm belonging \hs{2} to} \hs{2} q_A{\rm -brane} \\
{\rm otherwise}
\end{array},
\right.
\ena
and
\bea
g(t) = \left\{\begin{array}{ll}
\frac{1}{n-1} \ln \frac{\b}{\cosh[(n-1)\b(t-t_1)]} & :\s=+1, \\
\pm \b(t-t_1) & :\s=0, \\
\frac{1}{n-1} \ln \frac{\b}{\sinh[(n-1)\b|t-t_1|]} & :\s=-1,
\end{array}
\right.
\label{ints}
\ena
$t_A, t_1$ and $c$'s are integration constants which satisfy
\bea
&& c_0 = \sum_A \frac{q_A+1}{\Delta_A}\tilde c_A
-\frac{\sum_{\a=1}^p c_\a}{n-1}, \;\;
c_0' = -\frac{\sum_{\a=1}^p c_\a'}{n-1}, \;\;
\tilde c_\a = c_\a - \sum_A \frac{\c_{A}^{(\a)}-q_A-1}{\Delta_A}\tilde c_A, \nn
&& \tilde c_\phi = c_\phi + \sum_A \frac{(d-2)\e_A a_A}{\Delta_A}\tilde c_A.
\label{intconst1}
\ena
These must further obey the condition
\bea
\frac{1}{n-1}\left(\sum_{\a=1}^p c_\a\right)^2
+ \sum_{\a=1}^p c_\a^2 + \shalf c_\phi^2= n(n-1) \b^2.
\label{condconst}
\ena
So the solutions look sufficiently complicated that it may not be easy to
find the connection with the accelerating solution~\p{exsol1}.

Let us restrict these to a single S-brane in $d=11$ and set the field strength
to 0. Remember that the world-volume of $q$-branes lies in $(q+1)$-dimensional
space and not in time. For 11-dimensional supergravity, we have electric
SM2-branes (S2-branes in 11-dimensional supergravity), magnetic SM5-branes
and no dilaton $a_A=0, c_\phi=0$. Here we note that the relation between
$\tilde c_A$ and $c_\a$ in eq.~\p{oursol1} is derived under the assumption
that we have the independent field strengths $E_A$. In the absence of these,
we can disregard this relation and set $\tilde c_A$ to zero. We find that
the solution~\p{oursol} takes the form~\p{exsol1}-\p{factor} with
\bea
S(t)\equiv e^{-(n+2)(ct+c')/2(n-1)+ng(t)/2},
\ena
where we have set $c \equiv c_1=c_2=c_3$ and $c' \equiv c_1'=c_2'=c_3'$.
It then follows that our solutions reproduce the accelerating one~\p{exsol1}
if we further set $p=3, q_A=2, c=1, c'=0, t_1=0$ and $\s=-1$
(hyperbolic case in \p{ints}) with $\beta$ determined by eq.~\p{condconst}.

We note that there is a slight generalization in our solutions that allows
constant parameters $c$ and $c'$. We have also examined the possibility
if similar accelerating solutions can be obtained for flat and spherical
internal spaces. It turns out that neither the flat nor spherical internal
spaces do not give accelerating cosmologies; the condition for expansion
can be satisfied, but both cases give always decelerating universe.

\begin{figure}[htb]
\begin{minipage}{.45\linewidth}
\begin{center}
\setlength{\unitlength}{.7mm}
\begin{picture}(50,45)(10,5)
\includegraphics[width=6cm]{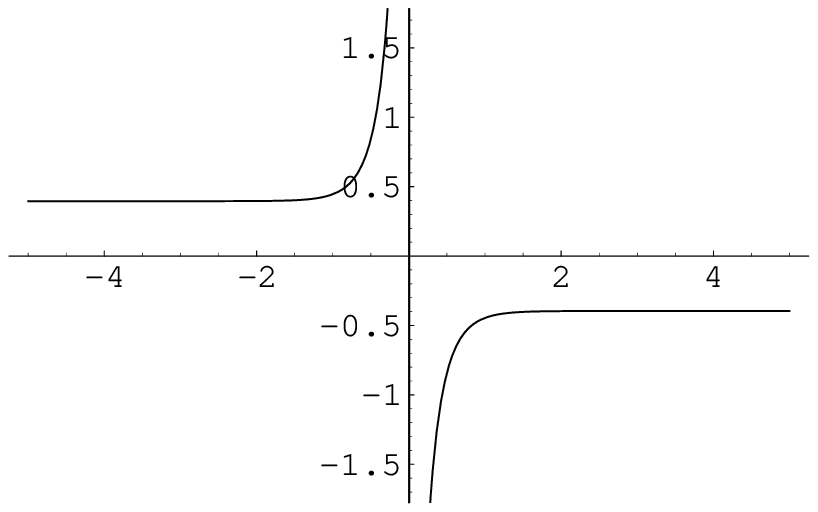}
\put(5,27){\footnotesize $t$}
\end{picture}
\caption{The lhs of eq.~\p{cond12}.}
\label{f1}
\end{center}
\end{minipage}
\hspace{10mm}
\begin{minipage}{.5\linewidth}
\begin{center}
\setlength{\unitlength}{.7mm}
\begin{picture}(50,45)(15,5)
\includegraphics[width=5cm]{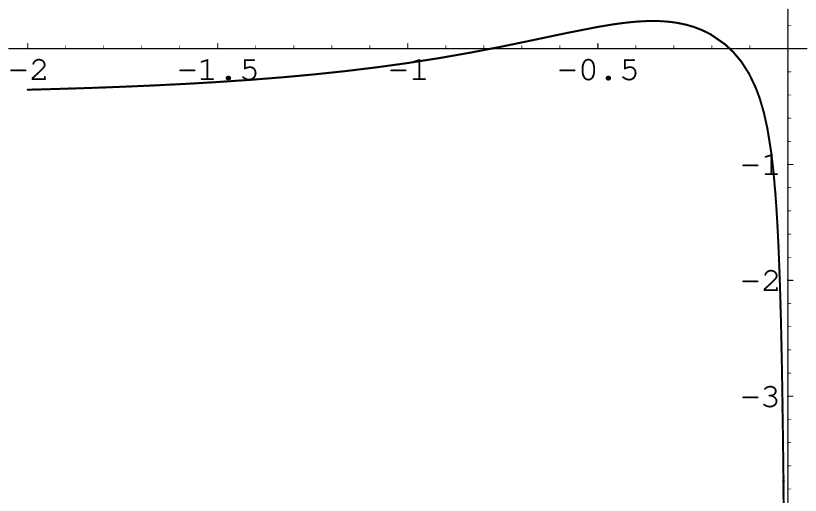}
\put(5,40){\footnotesize $t$}
\end{picture}
\caption{The lhs of eq.~\p{cond22}.}
\label{f2}
\end{center}
\end{minipage}
\end{figure}

We now show that our SM2-brane also gives 4-dimensional models of the
accelerating universe. We will find that here the flat internal space also
allows this kind of models. We choose $d=11, q_A =2, c\equiv c_1=c_2=c_3,
c' \equiv c_1'=c_2'=c_3'$. Our solutions~\p{oursol} then reduce to
\bea
ds_{11}^2 = [\cosh 3 c (t-t_A)]^{-7/6} e^{-7 g(t)+7c'/2} ds_E^2
+ [\cosh 3c(t-t_A)]^{1/3} e^{2g(t)-c'} d\Sigma_{7,\s}^2,
\label{coss1}
\ena
where the 4-dimensional part is given by
\bea
ds_E^2 = -[\cosh 3 c (t-t_A)]^{3/2} e^{21g(t)-9c'/2} dt^2
+ [\cosh 3c(t-t_A)]^{1/2} e^{7g(t)-3c'/2} d{\bf x}^2.
\ena
Comparing this solution with eqs.~\p{exsol1}-\p{factor}, we find that our
solutions have precisely the same form with $S(t)$ given by
\bea
S(t) = [\cosh 3c(t-t_A)]^{1/4} e^{7g(t)/2-3c'/4}.
\ena
We then define the time $\eta$ by eq.~\p{time} and examine if the
conditions for expansion~\p{cond1} and accelerated expansion~\p{cond2} are
satisfied. For $t_A=t_1=0$ and $\s=-1$ (hyperbolic space), we find
the condition \p{cond1} is
\bea
n_1(t) \equiv \frac{3}{4} \tanh(3ct)-\frac{\sqrt{21}}{4}\coth(3\sqrt{3/7}ct)>0,
\label{cond12}
\ena
and the condition~\p{cond2} gives
\bea
\frac{3}{2\sqrt{2}}\sqrt{\frac{1}{\cosh^2(3ct)}
+ \frac{1}{\sinh^2(3\sqrt{3/7}ct)}} - n_1(t) > 0.
\label{cond22}
\ena
The lhs of eqs.~\p{cond12} and \p{cond22} for $c=1$ are shown in
Figs.~\ref{f1} and \ref{f2}, respectively.
We see that there is a certain period of negative time that these conditions
are satisfied, exactly as the solution~\p{exsol1}. The period of the
accelerated expansion can be adjusted by changing the constant $c$.
Just as the accelerating solution~\p{exsol1}, the universe is decelerating as
$t\to -\infty\; (\eta \to 0)$ and $t\to 0$ from $t<0 \;(\eta \to \infty)$
\cite{TW}. The singularity at $t=0$ of the function $S(t)$ is at an
infinite proper time future for any event with $t<0$, and our universe
simply separates into two with $t<0$ and $t>0$.

\begin{figure}[htb]
\begin{minipage}{.45\linewidth}
\begin{center}
\setlength{\unitlength}{.7mm}
\begin{picture}(50,45)(10,5)
\includegraphics[width=5cm]{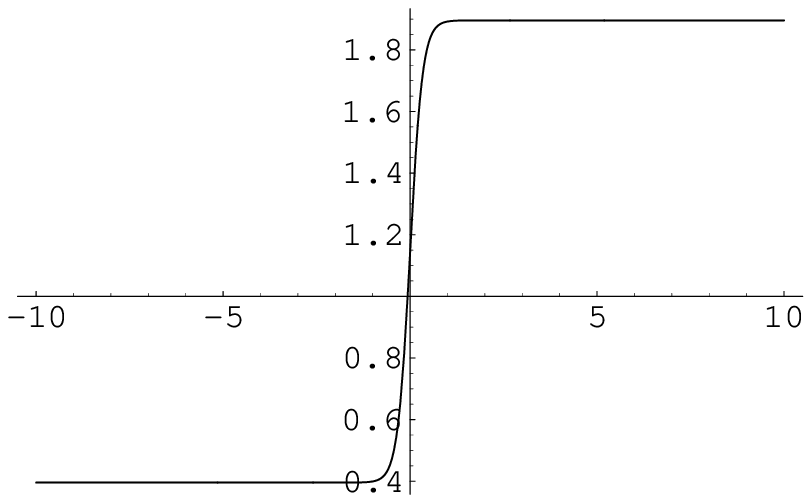}
\put(5,18){\footnotesize $t$}
\end{picture}
\caption{The lhs of eq.~\p{cond13}.}
\label{f3}
\end{center}
\end{minipage}
\hspace{10mm}
\begin{minipage}{.45\linewidth}
\begin{center}
\setlength{\unitlength}{.7mm}
\begin{picture}(50,45)(15,5)
\includegraphics[width=5cm]{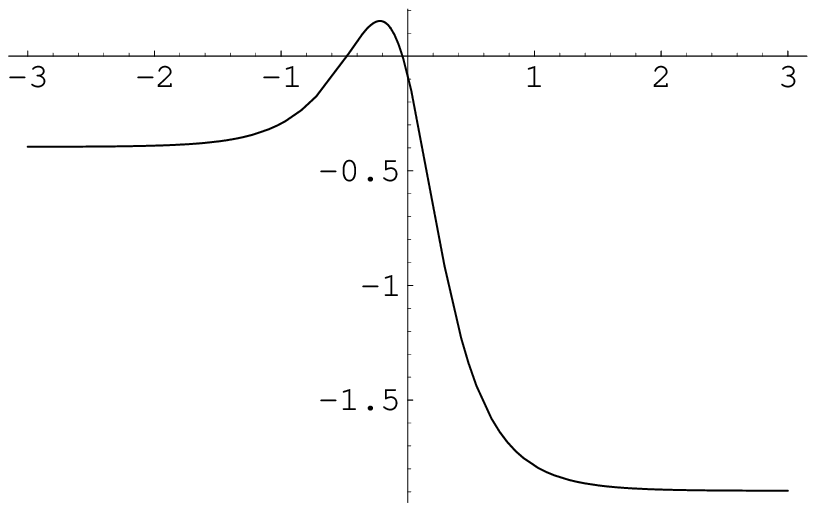}
\put(5,40){\footnotesize $t$}
\end{picture}
\caption{The lhs of eq.~\p{cond23}.}
\label{f4}
\end{center}
\end{minipage}
\end{figure}

If the internal space is chosen to be flat ($\s=0$), the conditions
\p{cond1} and \p{cond2} give
\bea
\label{cond13}
&& n_2(t) \equiv \frac{3}{4} \tanh(3ct) + \frac{\sqrt{21}}{4} > 0, \\
&& \frac{3}{2\sqrt{2}}\frac{1}{\cosh(3ct)}-n_2(t) > 0,
\label{cond23}
\ena
where we have chosen the plus sign in eq.~\p{ints} since minus sign cannot
give expanding universe. We find that these conditions can also be satisfied
for negative $t$ as shown in Figs.~\ref{f3} and \ref{f4}, which are the
results again for $c=1$.
Here we note that the universe is decelerating as $t\to -\infty\;
(\eta \to 0)$ and $t>0$. There is no singularity at $t=0$ and the time
$\eta$ start from 0 (at $t=-\infty$) to $\eta=\infty \; (t=\infty)$.
The accelerated expansion is realized for a certain period before $t=0$.

On the other hand, if we choose the internal space to be spherical ($\s=+1$),
we find that the conditions~\p{cond1} can be satisfied for negative $t$
but \p{cond2} cannot be satisfied for any value of the time.

As we have remarked above, the period of the accelerated expansion can
be changed by modifying the constant $c$ for hyperbolic and flat internal
spaces, but the expansion factor during the accelerated expansion
(the ratio of the scale factors at the starting time and ending time)
does not change. One typically obtains factor like 3, which is too small
to explain the horizon or flatness problems as a model of inflation at
the early universe. However, it is possible that solutions of large
amount of inflation can be found in this kind of models with suitable
modifications. Also the situation may change if we take into account of
quintessence field from matters. Another possibility is that the model
may be used for explaining the present accelerated expansion of the
universe. Details of the analysis on these problems will be reported
elsewhere~\cite{NO3}.

Though we have not examined other cases in 10-dimensional supergravities,
the only other S-brane solution that can give 4-dimensional universe is the
SD2-brane, which can be obtained from SM2-brane by dimensional reduction
and is expected to show similar behavior. However, it would be interesting
to further examine other possible solutions.

To summarize, we have shown that the accelerating solution~\p{exsol1} is a
special case of the solutions in ref.~\cite{NO1}. We have also shown that
the S-brane solutions can give interesting accelerating universe models
for the compact internal hyperbolic and flat spaces.
Other interesting time-dependent N-brane solutions have been found in
ref.~\cite{NO2}. It would be interesting to examine if this class of solutions
can give similar interesting cosmological models and also try to further extend
our analysis to other S-brane solutions.
We hope to discuss these problems elsewhere.

\section*{Acknowledgements}
We would like to thank P.K. Townsend and M.N.R. Wohlfarth for useful
correspondence, and J.X. Lu for valuable discussions. Most of the work was
done while the author was visiting the Interdisciplinary Center for
Theoretical Study at University of Science and Technology of China, whose
hospitality is gratefully acknowledged.
This work was also supported in part by Grants-in-Aid for Scientific Research
Nos. 12640270 and 02041.

\newcommand{\NP}[1]{Nucl.\ Phys.\ B\ {\bf #1}}
\newcommand{\PL}[1]{Phys.\ Lett.\ B\ {\bf #1}}
\newcommand{\CQG}[1]{Class.\ Quant.\ Grav.\ {\bf #1}}
\newcommand{\CMP}[1]{Comm.\ Math.\ Phys.\ {\bf #1}}
\newcommand{\IJMP}[1]{Int.\ Jour.\ Mod.\ Phys.\ {\bf #1}}
\newcommand{\JHEP}[1]{JHEP\ {\bf #1}}
\newcommand{\PR}[1]{Phys.\ Rev.\ D\ {\bf #1}}
\newcommand{\PRL}[1]{Phys.\ Rev.\ Lett.\ {\bf #1}}
\newcommand{\PRE}[1]{Phys.\ Rep.\ {\bf #1}}
\newcommand{\PTP}[1]{Prog.\ Theor.\ Phys.\ {\bf #1}}
\newcommand{\PTPS}[1]{Prog.\ Theor.\ Phys.\ Suppl.\ {\bf #1}}
\newcommand{\MPL}[1]{Mod.\ Phys.\ Lett.\ {\bf #1}}
\newcommand{\JP}[1]{Jour.\ Phys.\ {\bf #1}}



\begin{thebibliography}{99}
\bibitem{TW} P.K. Townsend and M.N.R. Wohlfarth, hep-th/0303097.
\bibitem{AR} A.G. Riess et al., Astrophys. J. {\bf 560} (2001) 49,
 astro-ph/0104455.
\bibitem{LB} A. Lewis and S. Bridle, \PR{66} (2002) 103511, astro-ph/0205436;\\
C.L. Bennett et al., astro-ph/0302207.
\bibitem{HW} C.M. Hull and N.P. Warner, \CQG{5} (1988) 1517.
\bibitem{T} P.K. Townsend, \JHEP{0111} (2001) 042, hep-th/0110072.
\bibitem{GH} G.W. Gibbons and C.M. Hull, hep-th/0111072.
\bibitem{GS} M. Gutperle and A. Strominger, \JHEP{0204} (2002) 018,
 hep-th/0202210.
\bibitem{CGG} C.M. Chen, D.V. Gal'tsov and M. Gutperle, \PR{66} (2002)
 024043, hep-th/0204071.
\bibitem{KMP} M. Kruczenski, R.C. Myers and A.W. Peet, \JHEP{0205} (2002)
 039, hep-th/0204144.
\bibitem{DK} N.S. Deger and A. Kaya, \JHEP{0207} (2002) 038, hep-th/0206057.
\bibitem{V} V.D. Ivashchuk, \CQG{20} (2003) 261, hep-th/0208101.
\bibitem{NO1} N. Ohta, \PL{558} (2003) 213, hep-th/0301095.
\bibitem{NO3} N. Ohta, hep-th/0304172 and paper in preparation.
\bibitem{NO2} N. Ohta, \PL{559} (2003) 270, hep-th/0302140.
\end{thebibliography}
\end{document}